\documentclass[sigconf, nonacm]{acmart}

\usepackage{array}
\usepackage[normalem]{ulem}
\useunder{\uline}{\ul}{}
\usepackage{longtable}
\usepackage{siunitx}
\AtBeginDocument{%
  \providecommand\BibTeX{{%
    \normalfont B\kern-0.5em{\scshape i\kern-0.25em b}\kern-0.8em\TeX}}}

\setcopyright{acmcopyright}
\copyrightyear{2018}
\acmYear{2018}
\acmDOI{XXXXXXX.XXXXXXX}

\acmConference[Conference acronym 'XX]{Make sure to enter the correct
  conference title from your rights confirmation emai}{June 03--05,
  2018}{Woodstock, NY}
\acmPrice{15.00}
\acmISBN{978-1-4503-XXXX-X/18/06}

\begin{document}

\title{Recy-ctronics: Designing Fully Recyclable Electronics With Varied Form Factors}


\author{Tingyu Cheng}
\affiliation{\institution{Georgia Institute of Technology}\city{Atlanta}\country{USA}}
\email{tcheng32@gatech.edu}

\author{Zhihan Zhang}
\affiliation{\institution{University of Washington}\city{Seattle}\country{USA}}
\email{zzhihan@cs.washington.edu}

\author{Han Huang}
\affiliation{\institution{Georgia Institute of Technology}\city{Atlanta}\country{USA}}
\email{hhuang449@gatech.edu}

\author{Yingting Gao}
\affiliation{\institution{Georgia Institute of Technology}\city{Atlanta}\country{USA}}
\email{ygao617@gatech.edu}

\author{Wei Sun}
\affiliation{\institution{Chinese Academy of Sciences}\city{Beijing}\country{China}}
\email{sunwei2017@iscas.ac.cn}

\author{Gregory D. Abowd}
\affiliation{\institution{Northeastern University}\city{Boston}\country{USA}}
\email{g.abowd@northeastern.edu}

\author{HyunJoo Oh}
\affiliation{\institution{Georgia Institute of Technology}\city{Atlanta}\country{USA}}
\email{hyunjoo.oh@gatech.edu }

\author{Josiah Hester}
\affiliation{\institution{Georgia Institute of Technology}\city{Atlanta}\country{USA}}
\email{josiah@gatech.edu}

\renewcommand{\shortauthors}{Cheng, et al.}

\begin{abstract}

For today's electronics manufacturing process, the emphasis on stable functionality, durability, and fixed physical forms is designed to ensure long-term usability. However, this focus on robustness and permanence complicates the disassembly and recycling processes, leading to significant environmental repercussions. In this paper, we present three approaches that leverage easily recyclable materials—specifically, polyvinyl alcohol (PVA) and liquid metal (LM)—alongside accessible manufacturing techniques to produce electronic components and systems with versatile form factors. Our work centers on the development of recyclable electronics through three methods: 1) creating sheet electronics by screen printing LM traces on PVA substrates; 2) developing foam-based electronics by immersing mechanically stirred PVA foam into an LM solution; and 3) fabricating recyclable electronic tubes by injecting LM into mold casted PVA tubes, which can then be woven into various structures. To further assess the sustainability of our proposed methods, we conducted a life cycle assessment (LCA) to evaluate the environmental impact of our recyclable electronics in comparison to their conventional counterparts.
\end{abstract}



\begin{teaserfigure}
  \includegraphics[width=\textwidth]{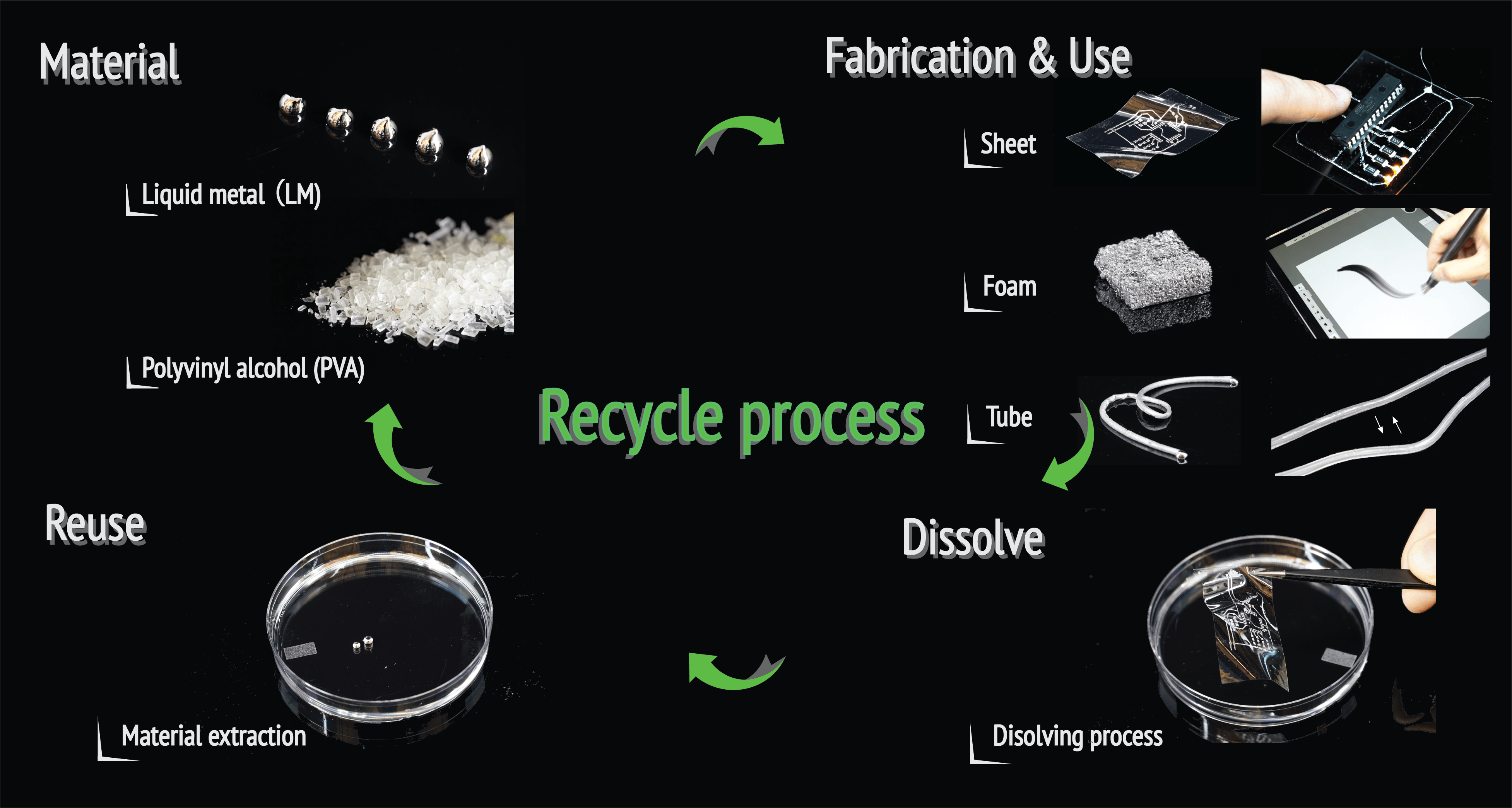}
  \caption{General recycling process for Recy-ctronics. 1: Materials preparation of PVA (substrate material) and liquid metal (electrode material). 2: Different recyclable interactive devices with various form factors (sheet, foam and tube). 3: The dissolving and separation process of the devices. 4: Material extraction for remaking new devices. 
 }~\label{figure:general}
  \label{fig:fig1}
\end{teaserfigure}

%
\maketitle

\section{Introduction}
The aspiration for 21st-century computing is its seamless integration into our everyday life (\textit{e.g.,} smart lock, smart home hub), a vision steadily coming to fruition with the proliferation of today's computational devices, serving users both locally and through the internet. However, the current landscape of connected devices (an average of 22 per U.S household in 2022) and global e-waste (53.6 million metric tons (Mt) by 2020) paints a bleak picture. The rapid expansion of IoT (Internet of Things) devices has resulted in environmental hazards surpassing our capacity to manage them sustainably. At the same time, as we venture into emerging fields like AI (Artificial Intelligence) and strive for fully immersive user experiences, the number of devices and subsequent waste will only escalate. Estimates indicate IoT devices could reach a trillion by 2035. This raises concerns about how to handle the disposal of these devices when they eventually become obsolete or malfunction. Addressing these questions is essential to building a sustainable future where technological advancement aligns with responsible environmental practices. 

Human-Computer Interaction (HCI) has a long history of prototyping various interactive devices, with a recent rise in promoting the development of sustainable ones. Traditionally, a significant portion of interactive devices rely on plastic housings (\textit{e.g.,} ABS, HDPE), conventional PCBs with FR4 and SMD components, and batteries containing harmful elements like lithium and lead. Improper processing or disposal of these components can pose serious environmental hazards, contributing to plastic waste, E-waste, and lithium battery pollution. Recognizing the environmental impact, researchers in HCI have been actively advocating for sustainable practices and reducing reliance on conventional materials in device manufacturing. For example, researchers have proposed transient and decomposable devices with more sustainable material options such as chitosan, beeswax to make wireless charged heaters or transient sensors \cite{cheng2023functional, song2022towards}. The focus of these works lies in transiency and biodegradability, prioritizing environmental sustainability over functionality or longevity.

In this work, we develop and integrate a set of materials, fabrication tools, and manufacturing processes to enable the accessible fabrication of recyclable electronic interactive devices. We would not only view the disintegration of the electronics as the end of their lifespan, but rather as marking the beginning of another circuit's physical and functional reconfiguration. Our aim is that these devices are designed for recycling efficiency--- unlike traditional electronics--- eliminating the need for sophisticated machinery or intricate chemical processes. To this end, we have chosen commercially available room-temperature liquid metals (RTLMs), particularly eutectic gallium-indium (EGaIn), for their conductive properties, and poly(vinyl alcohol) (PVA) as both substrate and encapsulation materials in the creation of recyclable electronics. 

Our innovations extend beyond traditional thin-sheet electronics, introducing three distinct form factors: sheets, foam, and tubes. We also introduce design strategies to customize devices' mechanical properties, ranging from rigid sheet electronics to flexible, compressible interactive foams, and highly stretchable tube sensors and actuators. Each form factor is tailored to meet different application needs while ensuring complete recyclability, showcasing our commitment to reducing electronic waste through innovative design and material selection.

The main contributions of this paper are:

\begin{itemize}
    \item Integration of PVA and LM to enable the creation of fully and easily recyclable electronics across three distinct form factors: sheet, foam and tube.
    \item Introduction of a formulation facilitating the production of three types of recyclable electronics with various properties: rigid, flexible, and stretchable.
    \item Development of a process to recycle three different forms of electronics with high recycling rates.
    \item A comparative life cycle assessment (LCA) case study for the recyclable sheet-based proximity sensor, validating the environmental benefits of the proposed approach and demonstrating ways to guide design decisions about sustainability without a full LCA.
\end{itemize}

\section{Related Work}
\subsection{Printed Electronics}
Printed electronics represents a cutting-edge technology that involves fabricating electronic devices and circuits through various printing methods, such as inkjet printing \cite{nayak2019review}, screen printing \cite{cheng2023swellsense} and 3D printing \cite{wang2018printed,valentine2017hybrid}. These printing techniques can also be combined with different pre/post-treatments for versatile use cases. For instance, conformally transfer printed heat-shrinkage film-based electronics can be applied to complex 3D surfaces \cite{rich2022developing}. Creating pre-strain or cutting patterns can enable stretchable electronics \cite{kim2008materials, talyan2023kirisense}. Printed electronics offer several advantages, including the creation of lightweight, mostly flexible or stretchable electronic components, circuits, or even systems. The manufacturing approach can also be more sustainable by favoring more on additive over subtractive methods.

Within the Human-Computer Interaction (HCI) community, rapid prototyping is a key theme for printed electronics research. One pioneering example is \textit{Instant Inkjet Circuits}, which introduced the use of commodity inkjet printers for quickly prototyping functional devices \cite{kawahara2013instant}. Building on this accessible and diverse design concept, researchers have utilized this technique to fabricate various sensors \cite{cheng2020silver, pourjafarian2019multi, yan2022fibercuit}, antennas \cite{li2016paperid}, displays \cite{olberding2014printscreen} and even heating-based actuators \cite{olberding2015foldio}. Moreover, researchers have demonstrated the scalability of printed electronics beyond the normal machine sizes. For example, \textit{Sprayable} introduced a spraying technique for prototyping large-scale sensors and displays \cite{wessely2020sprayable}, while \textit{Duco} introduced a system that leverages a hanging wall robot to draw large-scale circuits directly on everyday surfaces, using multiple tools and substrates.

\subsection{Recyclable Electronics}
With the remarkable advancements in fabricating diverse electronics, spanning various form factors, scales, and functionalities, the issue of disposing of these devices has become a pressing concern, yet not much effort has been devoted to this problem within the HCI community. Song et al. proposed a method to manufacture wireless heating interfaces that can decompose in backyard conditions within 60 days \cite{song2022towards}. \textit{Functional Destruction} is another recent work demonstrated three different approaches (PVA, beeswax and edible chocolate) to make electronics that are physically and functionally transient \cite{cheng2023functional}. Lazaro Vasquez et al. focused on mycelium, which is a fast-growing vegetative part of a fungus, to replace the plastic housing for traditional PCBs, but most of these work aims to make electronics with disposable or degradable manners instead of recyclable \cite{lazaro2019plastic}. The only two works, which are both from the same research group, try to emphasize recyclability for electronics. Arroyos et al. introduced an end-to-end digital fabrication process for creating a computer mouse with a biodegradable printed circuit board and case, utilizing materials like Jiva Soluboard and PVA 3D printing filament \cite{arroyos2022tale}, where the electronic components can be reused in this case. The other work expands the recyclability from the components to the substrate and the electrode material. Zhang et al. present vPCB, using vitrimer as the replacement for conventional circuit boards with a high recycling rate \cite{zhang_recyclable_2024}.

In contrast, other research fields such as materials science and electrical engineering have delved extensively into recycling electronics. Khrustalev et al., for instance, deviated from conventional thermoset materials like fiberglass and epoxy and instead used thermoplastic materials (3D printing PLA) as the substrate, showcasing how electronics can be fabricated and recycled efficiently \cite{khrustalev2022new}. \textit{3R} project also aims to design resilient, repairable, and recyclable electronics through a combination of novel biphasic printable LM composites and reversible tough block co-polymers \cite{tavakoli20223r}. LM, known for its excellent conductivity and intrinsic fluid nature, has been widely used as an electrode option for recyclable electronics. Teng et al. demonstrated two papers utilizing LM as the conductive material and PVA and beeswax as the substrate materials, respectively \cite{teng2019liquid, teng2023fully}. The beeswax approach stands out as it features a self-destructing-recycling process where the LM traces react with the sodium hydroxide within the device. These projects share a common goal of adapting straightforward recycling approaches that avoid complicated procedures or tooling while achieving high recyclability rates.

\subsection{Sustainable Computing}
The proliferation of miniaturized computational devices has brought up a range of environmental hazards, from plastic waste, E-waste to unsustainable power consumption. The call for sustainability echoes across industries, urging designers, researchers and makers to adopt more environmentally conscious materials and practices~\cite{mankoff2007environmental, lazaro2020introducing, brynjarsdottir2012sustainably, disalvo2010mapping, blevis2007sustainable, zhang_biodegradable_2024}. For material selection, even when eco-friendly options like PLA hold promise, it is always challenging to source from responsible manufacturers and find proper disposal avenues. Another critical area drawing from the use of traditional batteries~\cite{hester2017future,lucia2017intermittent}. While harvesting energy from various sources shows potential, it alone cannot ensure continuous device operation, prompting the need for solutions like energy buffering with capacitors to enable intermittent computing~\cite{de2020battery,bakar2022protean}. In the pursuit of sustainable alternatives, batteryless energy-harvesting devices have emerged as promising contenders, challenging the dominance of battery-powered counterparts with their eco-friendliness and reduced maintenance requirements~\cite{ferrese2015battery,de2022dips,bakar2021rehash, iyer_wind_2022}. Lastly, the concept of Life Cycle Assessment (LCA) is a cradle-to-grave method to study a device's environmental impact, which has guided researchers to evaluate whether a device is more or less sustainable in a more quantitative manner ~\cite{lazaro2020introducing}.

\begin{figure*}[h]
  \centering
  \includegraphics[width=1\linewidth]{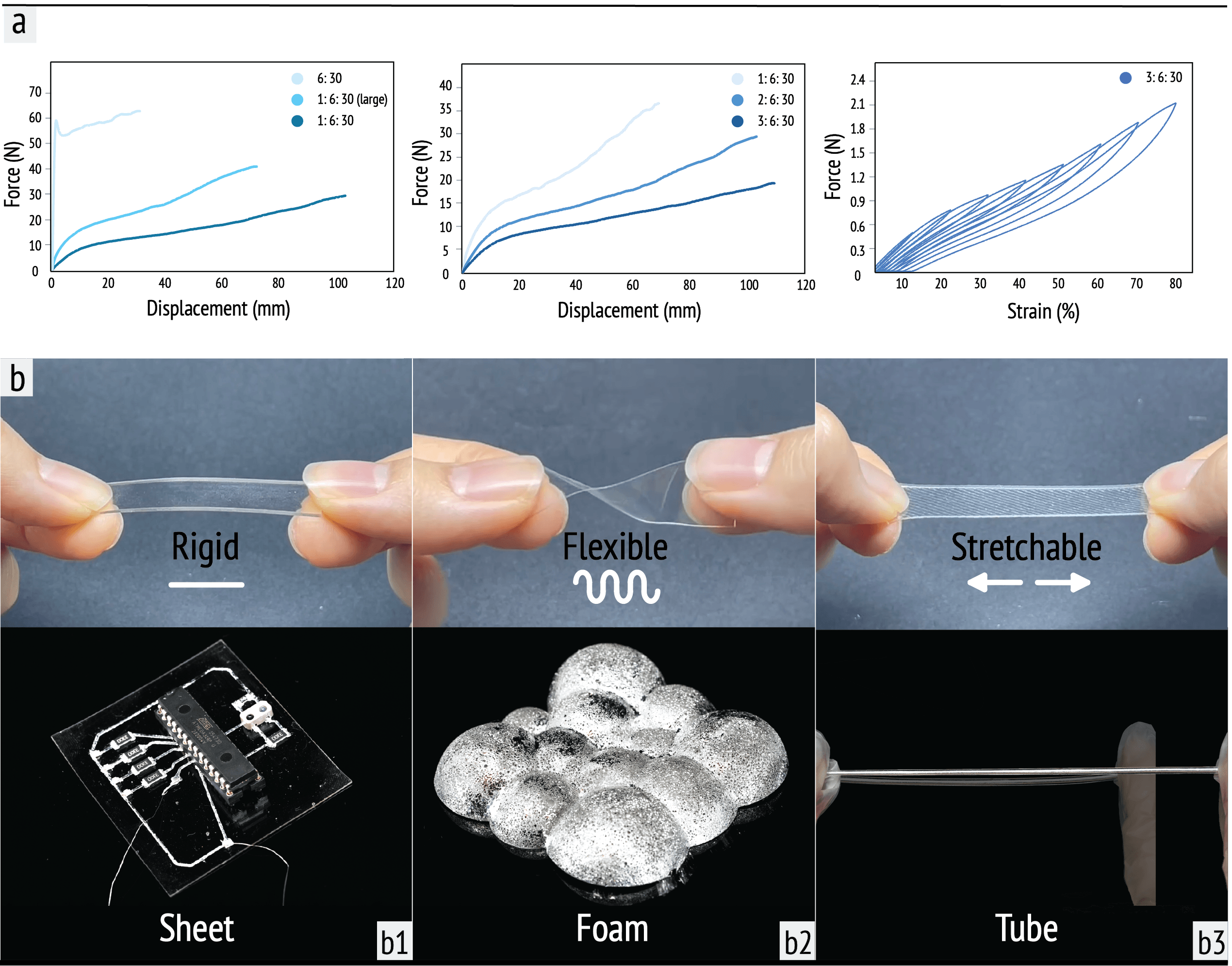}
  \caption{General design strategy. (a) Load versus strain curve under uniaxial deformation for different types of PVA and Cyclic loading to increasing strains as a function of strain. (b) Demonstration of PVA's mechanical flexibility (rigid, highly flexible, stretchable) for versatile types of devices (sheet, foam, tube). 
 }~\label{figure:testing1}
\end{figure*}

\section{General Recy-ctronics Design Rules}
Electronic devices have become pervasive in our daily lives, offering convenience while also giving rise to an inescapable tension between the rising quantity of these devices and the resulting waste they generate at the end of their life cycle. A central challenge in the realm of electronics recycling today stems from the prevailing focus on functionality, durability, and affordability during device creation, often at the expense of considering their sustainability at the end of their lifespan. Many devices were designed without prioritizing characteristics like "ease of repair" and "recyclability." In this section, we set three general design criteria for Recy-ctronics.

\begin{enumerate}
\item \textit{Recyclability over functionality:}
To enable "greener" electronics, we will prioritize materials selection and fabrication approaches that can ensure an easy, high-rate recycling process but meanwhile secure devices' functionality.

\item \textit{Versatile form factors and mechanical versatility:}
We aim to use recyclable materials to enable a wide range of electronics including form factors and mechanical properties.  

\item \textit{Highly accessible approaches and tools:} 
For Recy-ctronics, to support a broad spectrum of users, we expect our materials and tools selection can be low-cost, off-the-shelf parts, and easy to use.
\end{enumerate}

\begin{figure*}[h]
  \centering
  \includegraphics[width=1\linewidth]{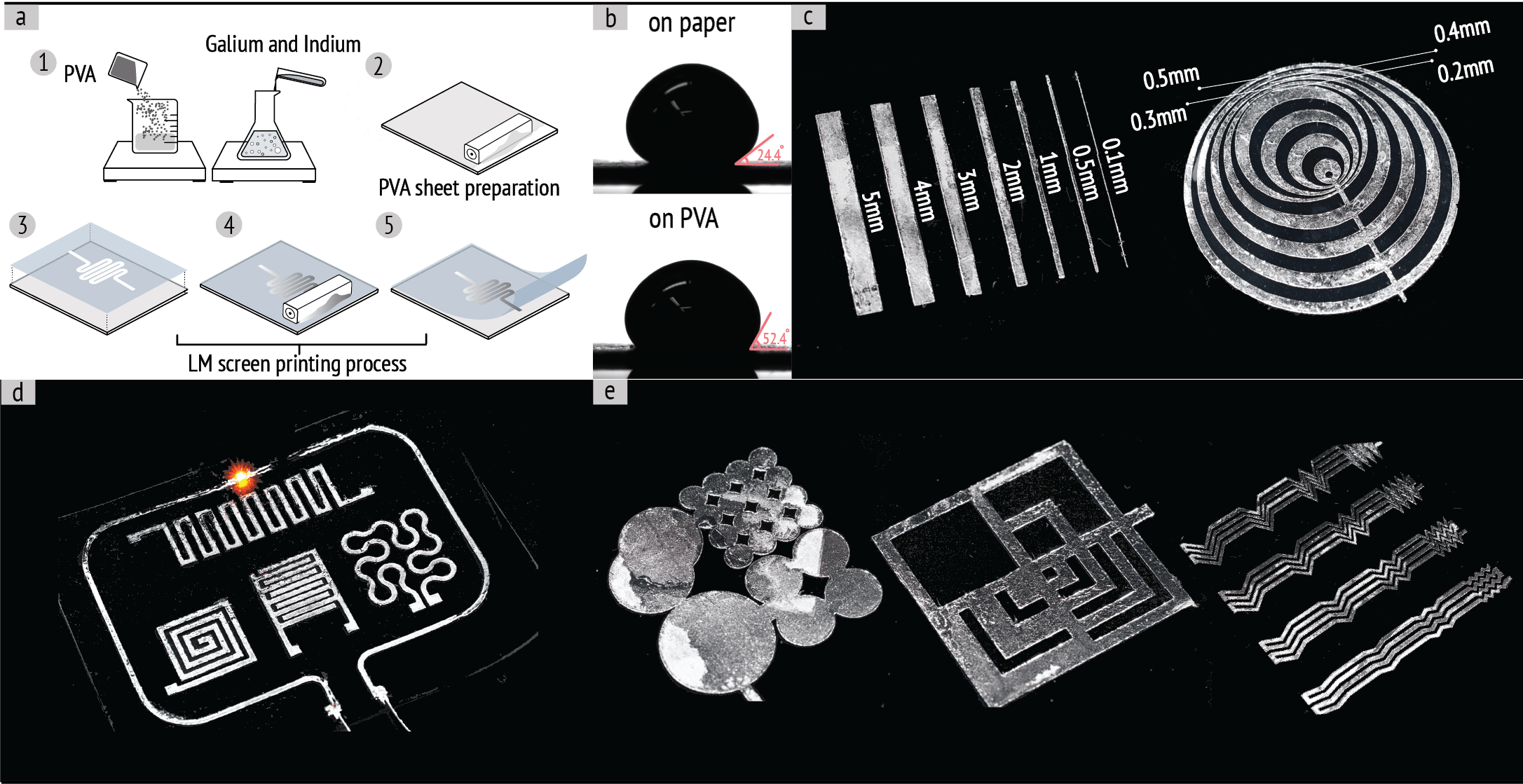}
  \caption{Sheet fabrication and basic characterization. (a) Fabrication process for making recyclable electronics with thin-sheet form factor. (b) Contact angle between LM and different substrates. (c) Resolution testing for trace width and gap width. (d,e) Different device and pattern design primitives. 
 }~\label{figure:sheet_overview}
\end{figure*}

\subsection{Recycling approach for Recy-ctronics}
The definition of electronics recycling nowadays is indeed broad, which is mainly referred to as e-waste recycling, pertaining to the retrieval of valuable materials and components from discarded electronic devices. However, when viewed through a more fundamental material lens, recycling processes can be further classified based on whether the items retain their original functionality. Recycling plastics as an example, approximately a decade ago, diverse strategies for recycling plastics were grouped into different categories: 1) primary, 2) secondary, 3) tertiary, and 4) energy recovery \cite{al2010valorization}. Primary recycling mainly involves reprocessing materials to generate substances with the same intended function. In contrast, secondary recycling yields materials repurposed for uses differing from the original plastic, often referred to as "downcycling". Tertiary recycling predominantly employs chemical processes, encompassing methods like hydrolysis, pyrolysis, hydrocracking, and gasification, to recover individual components or monomers. This approach presents opportunities for "upcycling" the original materials, leading to the production of potential higher-grade products. The fourth method, energy recovery, also termed quaternary recycling or incineration, involves harnessing heat as a form of energy recuperation. Recy-ctronics focuses mainly on primary and can be extended to secondary recycling processes. This means prioritizing electronics that can be recycled while retaining their initial functionality, with minimal or controlled levels of impurities. Of particular significance is our emphasis on facilitating recycling ease without any complicated procedures or chemical involvement, which can greatly enhance the electronics prototyping experience \cite{chiong2021integrating}.

\subsection{Versatile electronics from simple materials}
Electronics encompass a spectrum of components including semiconductors like silicon, metals like copper and gold, plastics, ceramics, glass, and rare earth elements. These materials are combined to form structures such as printed circuit boards (PCBs), connected by solder joints and powered by battery systems. Each component's characteristics, encompassing materials, shapes, and sizes, exhibit variations. In this paper, Recy-ctronics not only highlights ease of recycling and the quality of recycled materials, we also emphasize the flexibility, which we mean the capacity of materials to be adapted into diverse interactive devices. 

Polyvinyl Alcohol (PVA) stands as a synthetic polymer originating from vinyl acetate, synthesized via a distinct polymerization process. Unlike conventional polymerization routes, PVA emerges through the dissolution of polyvinyl acetate (PVAc) in alcohol like methanol, followed by treatment with an alkaline catalyst such as sodium hydroxide. This unique process grants PVA remarkable attributes including high water solubility, adeptness in forming films, and recyclability \cite{nagarkar2019polyvinyl}. These qualities have made PVA to be a favored material for transient electronics, serving as substrates and encapsulation mediums \cite{teng2019liquid}, which also made us choose PVA as the main material for Recy-ctronics. Besides the easy-to-dissolve nature of PVA, the mechanical versatility of PVA is also another key factor. As illustrated in \textcolor{blue}{Figure \ref{figure:testing1}} a, showcasing how we can tailor the mechanical property of PVA through simple adjustments in material composition and fabrication approaches. In general, for making PVA based substrates, a blend of glycerol/glycerin, de-ionized water, and PVA powder constitutes the primary components. Each component plays a different role in enabling diverse material properties. 

As shown in \textcolor{blue}{Figure \ref{figure:testing1}} a, we have conducted a series of tests on diverse PVA samples with varying compositions. As shown in the first graph of \textcolor{blue}{Figure \ref{figure:testing1}} a, if with the same recipe (glycerin: PVA: de-ionized = 1: 6: 30), PVA featuring a higher molecular weight (\textit{i.e.} M.W. 88,000-97,000) exhibits a higher elastic modulus, ultimate stress, and meanwhile demonstrates lower stretchability. Interestingly, within the same figure, if no glycerin is added, PVA behaves as a non-elastic material with clear yield stress and much higher Young's modulus, making it a great option as a rigid sheet material. Additionally, the overall stretchability can be augmented by increasing the proportion of glycerin, which is a typical plasticizer, during the PVA mixing process. As shown in the second graph of \textcolor{blue}{Figure \ref{figure:testing1}} a, it is clear when more glycerin is added to the system, more elastic and "softer" the material will become. Furthermore, on the right side of \textcolor{blue}{Figure \ref{figure:testing1}} a, we have carried out cyclic testing for PVA samples with a recipe of glycerin: PVA: de-ionized water ratio of 3: 6: 30. The sample underwent two cycles of increasing strain levels: 10\%, 20\%, 30\%, 40\%, 50\%, 60\%, 70\%, and 80\%. These tests revealed minimal plastic deformation of the sample during the process. All the testing specimens were prepared in dog-bone shape (ASTM D412A) and tested on a materials testing machine (Instron Universal Testing Systems 68SC-05) at a strain rate of 20mm/min. 

In this section, we aim to provide a simple strategy to easily tailor the mechanical property of PVA, which is a crucial step to enable a wide range of devices and applications. In general, as shown in \textcolor{blue}{Figure \ref{figure:testing1}} b, by varying the composition of PVA, one can achieve electronic materials from rigid to flexible to highly stretchable. This high tunability can also favor different forms of interactive devices we aim to make. For example, for making sheet-based electronics, the more rigid and non-elastic recipes can be used, while for fabricating foam-based electronics, we will mostly use the flexible recipes to enable the compressing behavior for foams, and for tubes, we prefer more stretchable recipes due to the typical stretchable application domains for this specific form factor.

\begin{figure*}[h]
  \centering
  \includegraphics[width=1\linewidth]{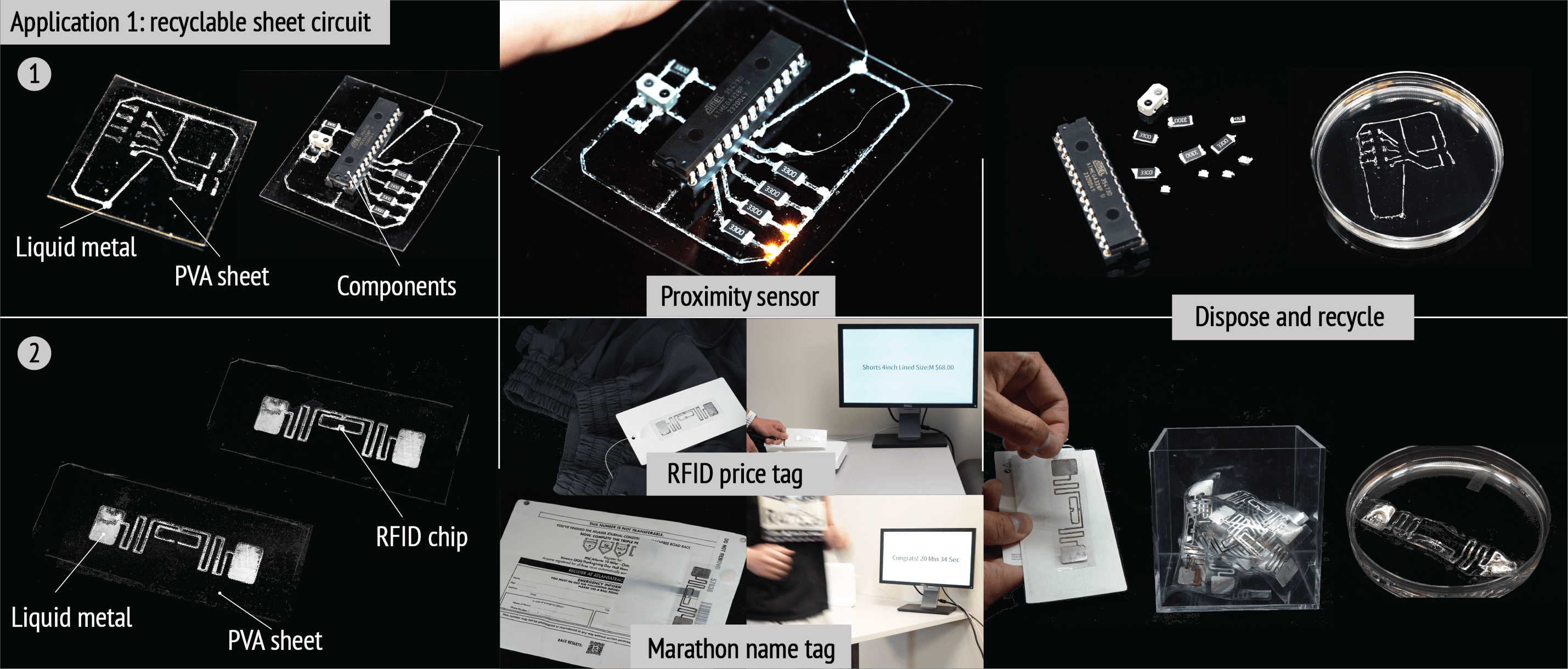}
  \caption{Sheet application. (1) Recyclable Sheet-based proximity sensor that responds to finger position. (2) Recyclable RFID tags used in two different application scenarios, including recyclable RFID clothing price tags and marathon RFID trackers. Both the proximity sensor and RFID tags are collected and recycled. 
 }~\label{figure:RFID}
\end{figure*}

\section{Sheet}
Even in today's electronics landscape, various form factors are present, yet planar geometry remains the main form factor (PCB board). Our initial exploration centers on using the Recy-ctronics approach to make some of today's electronics more sustainable. As shown in the initial step of \textcolor{blue}{Figure \ref{figure:sheet_overview}} a, both PVA and LM are prepared through mixing and stirring processes. In the formulation of PVA, a blend of glycerin, de-ionized water, and PVA powder constitutes the primary components. For crafting flexible and moderately stretchable thin sheets, a typical recipe involves a glycerin: PVA powder: de-ionized water ratio of 1: 6: 30. The LM involves the combination of pure Gallium and Indium in a weight ratio of 3: 1. Employing a magnetic stirrer (Thermo Scientific Cimarec stirring hot plate), PVA is subjected to mixing at 80°C for approximately 2 hours, while LM undergoes a 160°C overnight stirring process. Once both materials attain the desired state, a film applicator (GLTL four-sided coating applicator) is utilized to form a PVA film. This film is allowed to cure either at room temperature overnight or in an oven for 4 hours at 60°C depending on thickness. Subsequently, a stencil (Selizo low tack transfer paper) is applied to the PVA sheet, onto which LM is evenly brushed with the assistance of a rubber roller. The stencil is removed post circuit completion (shown in \textcolor{blue}{Figure \ref{figure:sheet_overview}} a). We conducted preliminary tests to examine the printing resolution for trace and gap widths, achieving a minimum trace width of 0.1mm and a smallest gap width of 0.2mm (\textcolor{blue}{Figure \ref{figure:sheet_overview}} c). Moreover, we present several printed electronics device design- LED, inductor, capacitor and heater pattern (\textcolor{blue}{Figure \ref{figure:sheet_overview}} d). We have also explored some pattern designs including interconnected circles, squares and waves with different density—showcasing the design and fabrication potential inherent in the utilization of PVA and LM (\textcolor{blue}{Figure \ref{figure:sheet_overview}}  e). It also shows great adhesion and intrinsic mechanical compatibility between PVA and the room-temperature LM. Usually, because of the high surface energy of LM, it barely gets attached to other surfaces. As shown in \textcolor{blue}{Figure \ref{figure:sheet_overview}} b, the contact angle of PVA and LM is 52.4$\pm$4.2$^\circ$ (ramé-hart Contact Angle Goniometer), indicating a relatively good adhesive between PVA and LM, and as a comparison, the contact angle between LM and the copy paper is also measured, which is much smaller (24.4$\pm$2.5$^\circ$). With this great adhesion between LM and PVA sheet, we have also achieved a sheet resistance of 0.013$\pm$0.002 $\Omega/sq$, where 20 samples with 40cm by 0.5cm dimension were tested (B\&k Precision 891 LCR Meter).

\subsection{Application 1: Recyclable Sheet Electronics--- from multiple components to minimal components}

Our primary goal is to enable the transformation of the current Printed Circuit Boards (PCBs) into easily recyclable alternatives using PVA and LM. In this section, we highlight two distinct circuit designs to demonstrate the versatility of Recy-ctronics in prototyping electronics with varying component complexities—from those with many components to those with minimal components.

As shown in the top row of \textcolor{blue}{Figure \ref{figure:RFID}}, we first introduce a proximity sensor circuit featuring an OPB733TR, which integrates both an infrared LED and an NPN phototransistor. This sensor activates additional LEDs as a hand or object passes over it, with the lights dimming once the object is removed. This interaction is managed by a standard Microcontroller Unit (MCU), specifically an ATMEGA328P-PU. To enhance compatibility with the sheet circuit, a socket is added for the MCU, ensuring a flat connection between the MCU and the sheet substrate. This assembly is bonded using commercial silver epoxy. This particular circuit design incorporates twelve electronic components in total, including the phototransistor, the MCU, four Surface-Mount Device (SMD) LEDs, and six SMD resistors. The PVA substrate used in this specific circuit design has a dimension of 5.5cm x 4.5cm x 0.5mm, made from a mixture excluding glycerin (PVA: de-ionized water = 1: 5). Also, all components are designed for easy disassembly and reuse, and the substrate dissolves entirely in water within ten minutes, enabled by the PVA circuit's recyclable nature (shown on the right side of \textcolor{blue}{Figure \ref{figure:RFID}}).

Besides the proximity sensing circuit, which can exemplify most of today's circuit, there are also electronics that have minimal to no components but serve a crucial role in our everyday life. Passive RFID technology is one typical example, which has gained widespread adoption in various everyday scenarios (such as key fobs, door access, chipless payments, etc.), primarily due to its technological benefits. Firstly, it operates without the need for a battery. When an RFID reader is in proximity, the tag becomes instantly powered, enabling it to transmit the data stored within it back to the reader. Secondly, its design is notably slim, enabling seamless integration of the RFID tag into thin and flat devices (like clothing price tags, apartment key fobs, etc.). Lastly, the production cost of the RFID tag is very economical, facilitating large-scale manufacturing. These benefits make RFID not only a promising commercial product in almost all industry branches, also draws attention in using RFID for interaction, including tracking 20 objects and identifying their movements \cite{li2015idsense}, utilising RFID to build interaction with smart devices \cite{li2016paperid}.  

However, the widespread adoption of RFID tags has also brought about concerns regarding their end-of-life waste management. Recent research highlights that selected logistics centers alone contribute to an annual generation of 5.7 tons of e-waste from discarded radio frequency identifiers on received pallets, with this waste containing 139 kg of metal components. Our primary objective is to address this issue by introducing easily recyclable RFID tags, specifically targeting two key scenarios. Illustrated in the second row of \textcolor{blue}{Figure \ref{figure:RFID}}, we have affixed recyclable ultra-high-frequency (UHF) RFID tags to clothing price tags and marathon name tags. After these RFID tags have fulfilled their purpose, users can conveniently deposit them into designated collection bins (depicted in the third column of \textcolor{blue}{Figure \ref{figure:RFID}}), initiating a process of dissolution and subsequent recycling.

As shown in \textcolor{blue}{Figure \ref{figure:RFID}}, we fabricated the RFID tags utilizing PVA as the substrate, overlaying it with an LM pattern. Employing the Impinj Monza 4QT chip, which can either be retrieved prior to RFID device dissolution or recycled subsequently, added to the fabrication process. Our measurement approach employed an ultra-high-frequency (UHF) RF reader, specifically the ImpinJ Speedway R420 RFID reader paired with the Laird S9028PCR antenna, encompassing the radio frequency range spanning 902MHz to 958MHz. Leveraging the Octane SDK, we established a comprehensive system that not only reads UHF RFID tags but also furnishes pertinent feedback. As presented in the central column of \textcolor{blue}{Figure \ref{figure:RFID}} b, we conducted replicative scenarios involving the utilization of our recyclable RFID tags in shopping and marathon runner information tracking. Remarkably, in both scenarios, the recyclable RFIDs consistently exhibited robust detectability by our reader systems. Also, the fabrication process is similar as the proximity sensor example. However, considering most of today's commercial RFID tags are in a more flexible form, so we fabricated the RFID on a more flexible substrate recipe (glycerin: PVA: de-ionized water = 1: 6: 30), which can be more conformally attached to the price tag or marathon name tag.

\begin{figure*}[h]
  \centering
  \includegraphics[width=1\linewidth]{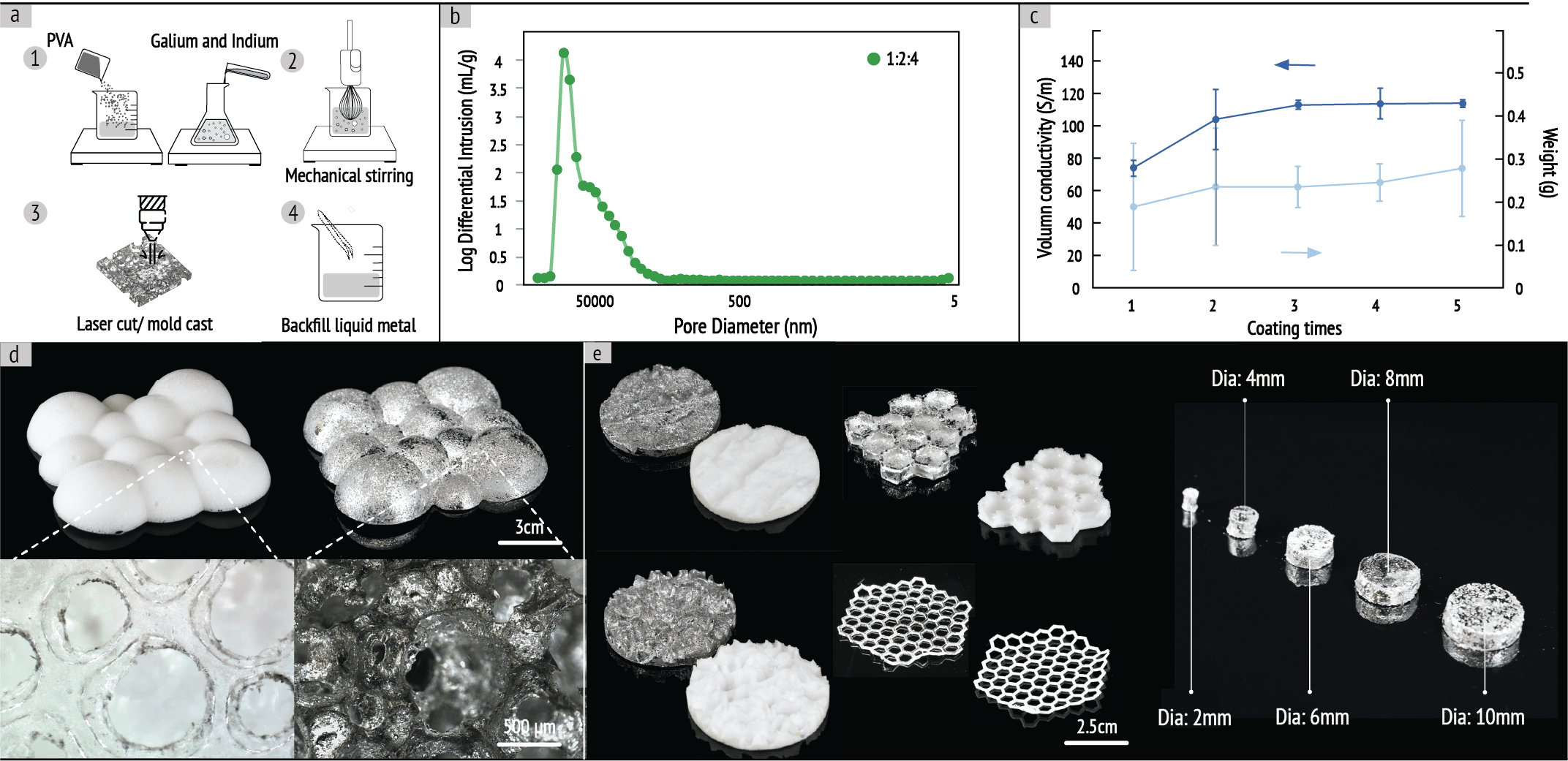}
  \caption{Foam fabrication and basic characterization. (a) Fabrication process for making recyclable foam electronics. (b) The pore diameter for the 1:2:4 foam recipe. (c) The conductivity testing for the interactive foam. (d,e) Different types and dimensions of recyclable interactive foams through mold casting or laser cutting. 
 }~\label{figure:foam1}
\end{figure*}

\section{Foam} Besides sheet form factors, which are more of a conventional form factor for printed circuits, foam, with its inherent porous composition and responsive elasticity, has garnered significant attention in HCI, offering diverse avenues for crafting interactive devices. Nakamaru et al. showcase foam devices coated with conductive ink, enabling volumetric alterations for detecting a spectrum of user inputs: compression, bending, torsion, and even shearing \cite{nakamaru2017foamsense}. In this section, our study aims to use PVA and LM to enable fully recyclable interactive foam structures.

As shown in \textcolor{blue}{Figure \ref{figure:foam1}} a, the fabrication process starts by preparing both the PVA material and LM. The sequential steps are similar to the ones described in the sheet fabrication section. However, a major difference arises in the formulation for PVA and how we coat the LM. To ensure the foam's stability during the curing phases of PVA, we reduce the use of water content coupled with an increased glycerin proportion. The main recipe we adopt in this study covers a blend of glycerin: PVA: de-ionized water in a ratio of 1: 2: 4. Once the PVA solution is prepared, a mechanical stirrer operating at approximately 2000rpm for 6 minutes serves to transform the PVA solution into a spongy and porous configuration. It's noteworthy that we had also tested adding nano-cellulose fibers (NCF) and surface surfactants (such as Sodium dodecyl sulfate) which can effectively maintain the structure of the recyclable foam during the curing process, meaning no foam bubbles will collapse during the curing step but will complicate the material synthesizing and also the subsequent recycling process. After the complete drying process of the PVA foam, which is thickness dependent (typically spanning 1 day at room temperature for a 2mm foam), laser cutting can be employed to shape the foam into versatile geometries. Concluding the process, as shown in \textcolor{blue}{Figure \ref{figure:foam1}} a, step 4, the foam is immersed in LM, with an additional vacuum treatment to back-fill the LM into the foam's internal voids. \textcolor{blue}{Figure \ref{figure:foam1}} b and c show two testing results, including the pore concentration for the 1: 2: 4 foam, which is mainly around 55000nm (AutoPore 9520 Mercury intrusion porosimeter) and also the conductivity and weight changes across different back-fill times. It is worth noting that as we back-fill more times, LM will flow and attach to more voids within the PVA foam, making the foam more conductive. For shaping the foams into different geometries, not only laser cutting, foams can also be mold-casted, shown in \textcolor{blue}{Figure \ref{figure:foam1}} d and e, we have introduced different recyclable interactive foams with various sizes, geometries or even surface textures.

\subsection{Foam-based Electronics Primitives}

\begin{figure*}[h]
  \centering
  \includegraphics[width=1\linewidth]{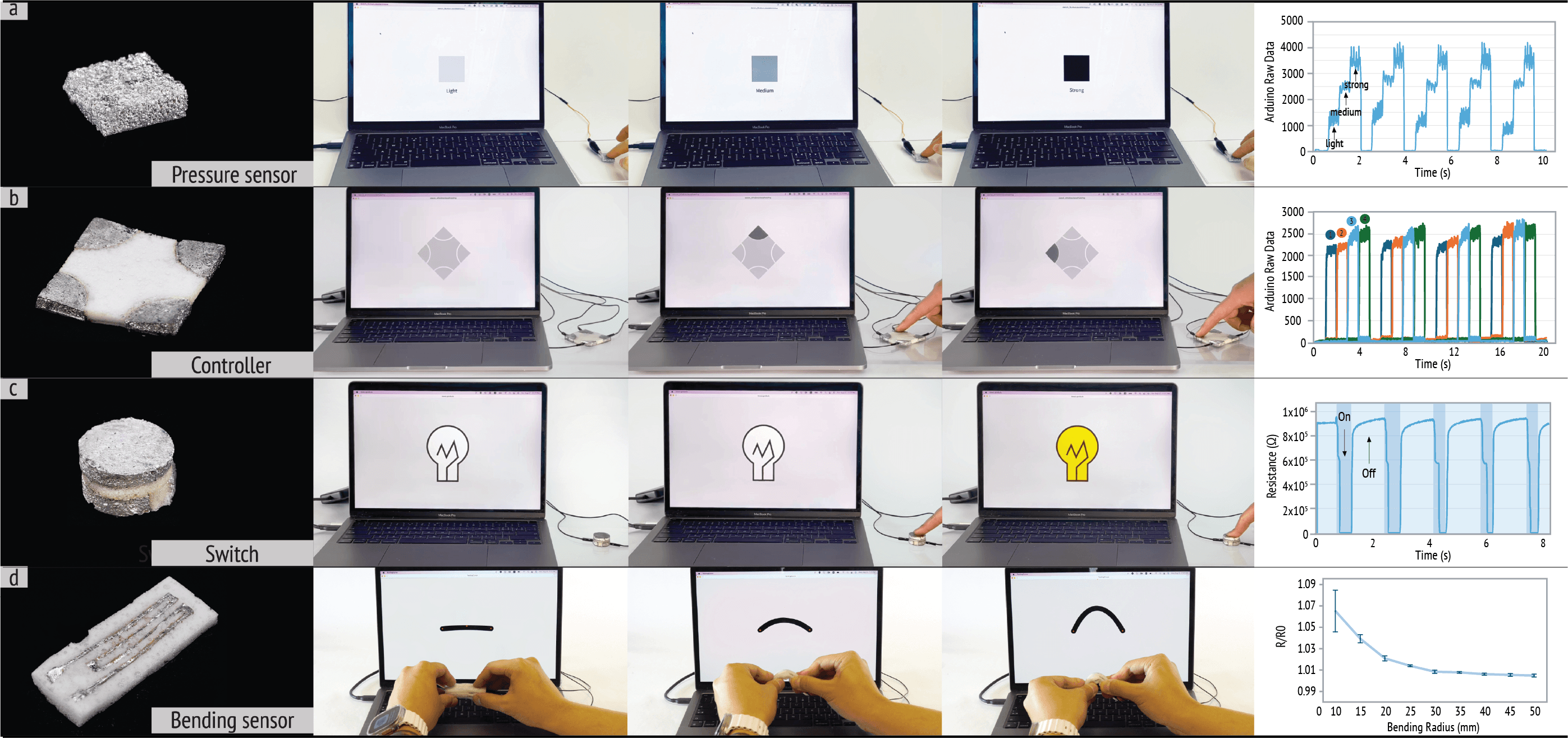}
  \caption{Foam sensing primitives. (a, b) Capacitive sensing based pressure sensor and controller and (c, d) Resistive sensing based mechanical contact switch and bending sensor. 
 }~\label{figure:foam_primiteive}
\end{figure*}

Foam, with its unique bouncy property, can enable unique multifunctional sensing capabilities. Here, we present four innovative conductive foam designs. We initiated our exploration by showcasing a basic square foam configuration as a pressure-sensitive unit, which can detect different levels of user-applied pressures (\textcolor{blue}{Figure \ref{figure:foam_primiteive}} a). This foam pressure sensor is able to distinguish between light, moderate, and firm touches. Different than this fully LM-soaked pressure sensor, in \textcolor{blue}{Figure \ref{figure:foam_primiteive}} b, we introduce a selectively conductive foam. Here, we utilized a laser cutter to cut the four corners of the foam, selectively immerse these units into LM to make them conductive, and then reattach them back to the original piece by using PVA solution as glue. This device now functions as a controller, offering four distinct input options. Both the pressure sensor and controller implementations rely on self-capacitance mechanism and the Arduino CapacitanceSensor Library, with data acquisition handled by an Arduino UNO (shown in \textcolor{blue}{Figure \ref{figure:foam_primiteive}} a and b), and 1M$\Omega$ resistor was used in the RC circuit.

Expanding beyond capacitive sensing, we showcase a mechanical-contact-based switch in \textcolor{blue}{Figure \ref{figure:foam_primiteive}} c, where we present a three-layer foam structure comprising a fully conductive top layer, an insulating middle layer, and a bottom layer housing two separated conductive components. Utilizing the foam's exceptional compressibility, applying pressure to the top conductive layer will make contact with the two lower terminal layers, which further completes a closed-loop circuit. This configuration has the potential to serve as a functional light switch. While bending sensors have been frequently explored, it becomes challenging to use highly conductive LM, which has minimal resistance variance during bending. In \textcolor{blue}{Figure \ref{figure:foam_primiteive}} d, thanks to the laser cutting approach, we try to overcome this challenge by laser-cutting a strain-gauge-like pattern to amplify resistance changes along the bending direction. Both the mechanical-contact switch and the enhanced bending sensor (\textcolor{blue}{Figure \ref{figure:foam_primiteive}} c and \textcolor{blue}{Figure \ref{figure:foam_primiteive}} d) are implemented using Arduino UNO and voltage-dividing mechanisms. We have also included the characterization data for both sensors on the right side of \textcolor{blue}{Figure \ref{figure:foam_primiteive}} c and d.

\subsection{Application 2: Recyclable Interactive Pen Tip}

\begin{figure*}[h]
  \centering
  \includegraphics[width=1\linewidth]{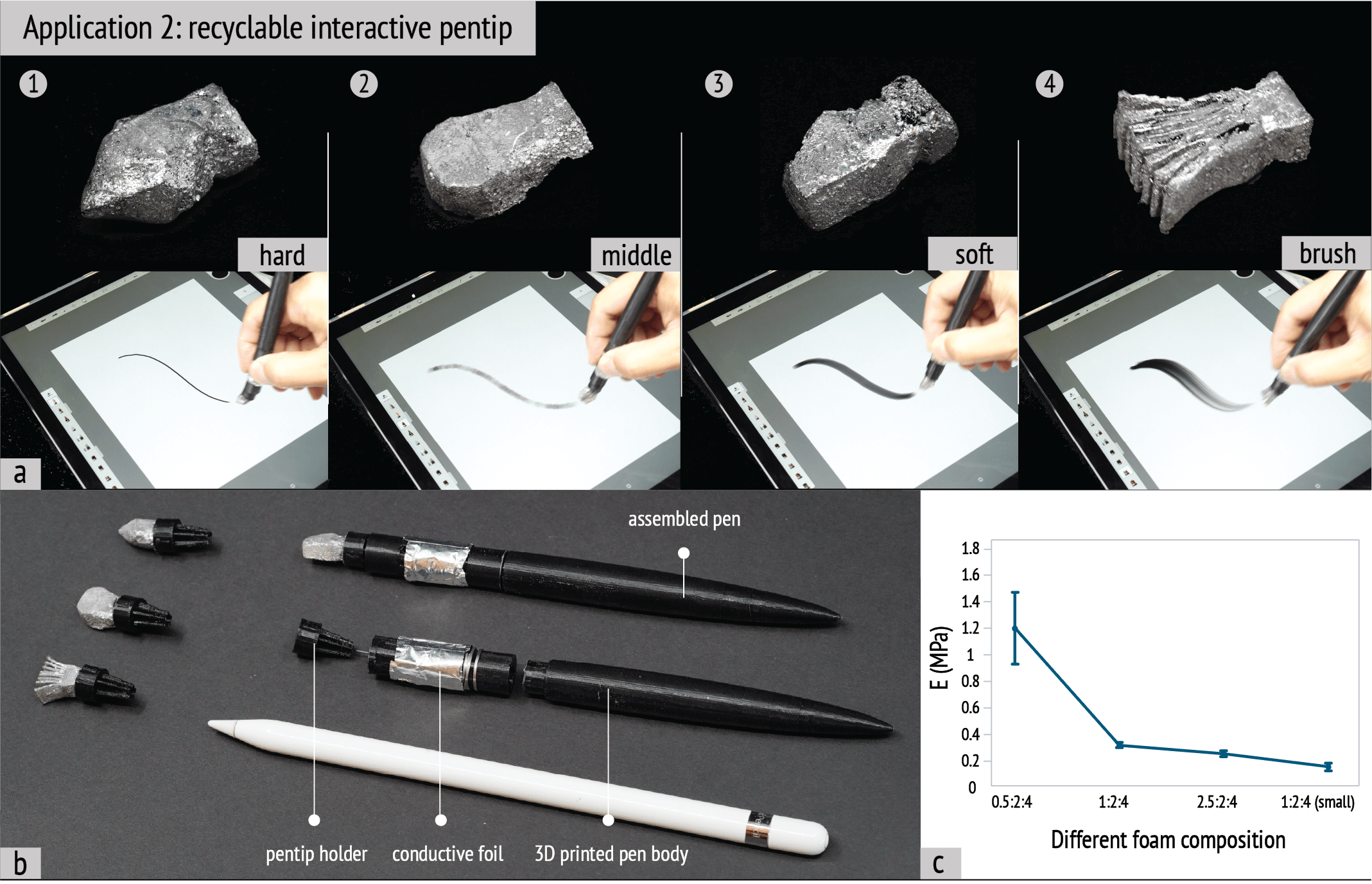}
  \caption{Recyclable interactive pen tip application. (a) From 1-4, we showcase recyclable interactive pen tips with different hardness and tactile feelings, from hard, medium, soft to brush feelings. (b) The disassembly look of the interactive pen for mobile devices with replaceable pen tips. (c) Young's modules for the foams with different material compositions.
 }~\label{figure:foam2}
\end{figure*}

Foam has many distinctive qualities, encompassing its softness, cushioning attributes, and adaptability to the contours of objects or bodies. One of the most notable characteristics when interacting with foam-based devices is the bouncy tactile feedback. Applying this mechanical property, we present an interactive pen featuring four distinct interactive and recyclable pen tips. We aim to address the existing gap in current drawing software, which typically offers a wide array of pen tools, while conventional interactive pens, such as the Apple Pencil, only provide rigid tips that do not replicate the tactile sensations associated with specific tools. As illustrated in \textcolor{blue}{Figure \ref{figure:foam2}} a 1-4, we demonstrate four pen tips with varying foam properties designed to mimic the appearance of different drawing tools, such as a pencil, ballpoint pen, and brush. Also, we employed various formulations to create PVA foams with adjustable levels of hardness, enabling these four interactive pen tips to provide users with sensations closely resembling real pen tools. As shown in \textcolor{blue}{Figure \ref{figure:foam2}} b, these pen tips can be affixed to 3D-printed pens, with a conductive foil and wire used to establish a capacitive coupling between the pen tips and the user's hand. The formulations utilized for creating hard, medium, soft and brush pen tips are as follows: glycerin: PVA: de-ionized water ratios of 0.5: 2: 4, 1: 2: 4, and 2.5: 2: 4, and 1: 2: 4 (small PVA molecules) respectively. We have also included the young's modulus of these four different foams in \textcolor{blue}{Figure \ref{figure:foam2}} c, showing their distinct hardness. It is important to note that only the pen tips are fully recyclable in this application, while the pen holder and the pen's body are 3D printed using black PLA materials.

\section{Tube}
\begin{figure*}[h]
  \centering
  \includegraphics[width=1\linewidth]{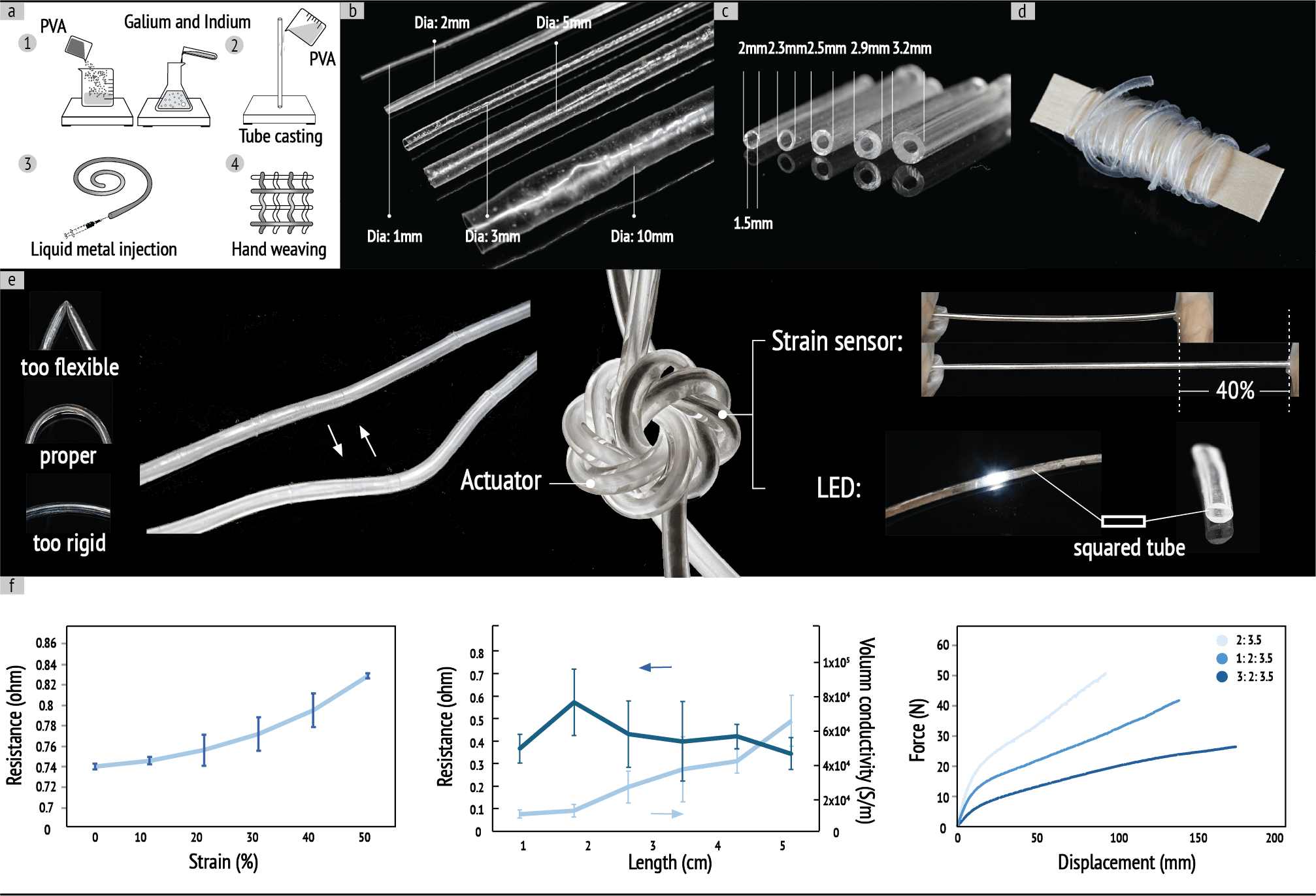}
  \caption{Tube fabrication and basic characterization. (a) Fabrication process for making recyclable tube electronics. (b-d) Recyclable tube with different diameters, wall thickness and length. (e) Three types of tube-based devices, including LED, sensor and actuator. (f) Basic electrical and mechanical testing results for tubes.
 }~\label{figure:tube}
\end{figure*}

Besides sheet and foam, Recy-ctronics is introducing a third form factor: tube. Tubes, tailored by their dimensions and diameter-to-length ratios, are versatile in applications, allowing for the creation of bending actuators \cite{luo2022digital} or integration into textiles for sensing purposes \cite{luo2021knitui}. This section focuses on utilizing PVA and LM-based tubes to develop various types of hand-weaving recyclable devices. The method for fabricating these recyclable tubes begins with the preparation of PVA solution and LM, similar to the processes for sheet and foam. We exemplify three different recipes for making different types of tubes in the third graph of \textcolor{blue}{Figure \ref{figure:tube}} f, where by varying the mix ratios, it is possible to produce tubes ranging from rigid and flexible to highly stretchable. While for most of the tubes used in this paper, a mixture ratio of glycerin: PVA: de-ionized water = 3: 2: 3.5 is preferred for achieving high elasticity in this paper.

The fabrication technique for tubes is straightforward, as illustrated in \textcolor{blue}{Figure \ref{figure:tube}} a. A simple casting method involves dipping a carbon fiber rod into the PVA solution. The process is complete once the PVA tube has cured and the curing time is similar to curing PVA sheet which is also thickness dependent. As demonstrated in \textcolor{blue}{Figure \ref{figure:tube}} b, tubes of varying diameters can be produced by using carbon fiber rods of different sizes. Additionally, as shown in \textcolor{blue}{Figure \ref{figure:tube}} c, applying multiple layers of PVA allows for the creation of tubes with various wall thicknesses. While this tube casting method is limited to producing tubes up to one meter in length, corresponding to the length of carbon fiber rod that is available, longer tubes can be fabricated by joining multiple tubes end-to-end. This is shown in \textcolor{blue}{Figure \ref{figure:tube}} d, where a 3-meter tube was assembled by connecting several tubes together using PVA as an adhesive.

In this paper, we present two types of tubes for making interactive devices: hollow tubes and LM-filled tubes. The stretchable LM-filled tubes can be utilized as strain sensors. Through experimentation, we detail the process by stretching these tubes to a 50\% strain results in a resistance increase from approximately 0.74$\Omega$ to 0.82$\Omega$ (shown in first graph of \textcolor{blue}{Figure \ref{figure:tube}} f). Also, we explore the possibility of using LM-filled tubes for connecting LEDs. Unlike the circular tubes, we chose to
use squared tubes that can align more compatibly with the geometry of standard SMD LEDs (as illustrated on the right side of \textcolor{blue}{Figure \ref{figure:tube}} e). Beyond sensors and LEDs, which both rely on LM-filled tubes, we also showcased the application of hollow PVA tubes in the development of recyclable pneumatic actuators. To enhance the functionality of these tube actuators, we add a constraining layer by using fishing lines, which will limit the axial tube deformation and only the longitudinal deformation is allowed. Besides, on the left side of \textcolor{blue}{Figure \ref{figure:tube}} e, we have investigated the design consideration for using tubes for fabricating devices. Tubes that are overly flexible with thin walls may easily get sharp bends, obstructing the flow of LM or air. Conversely, tubes that are too rigid pose difficulties in bending and integration into textiles, limiting their applicability in wearable technology and smart fabrics.

\begin{figure*}[h]
  \centering
  \includegraphics[width=1\linewidth]{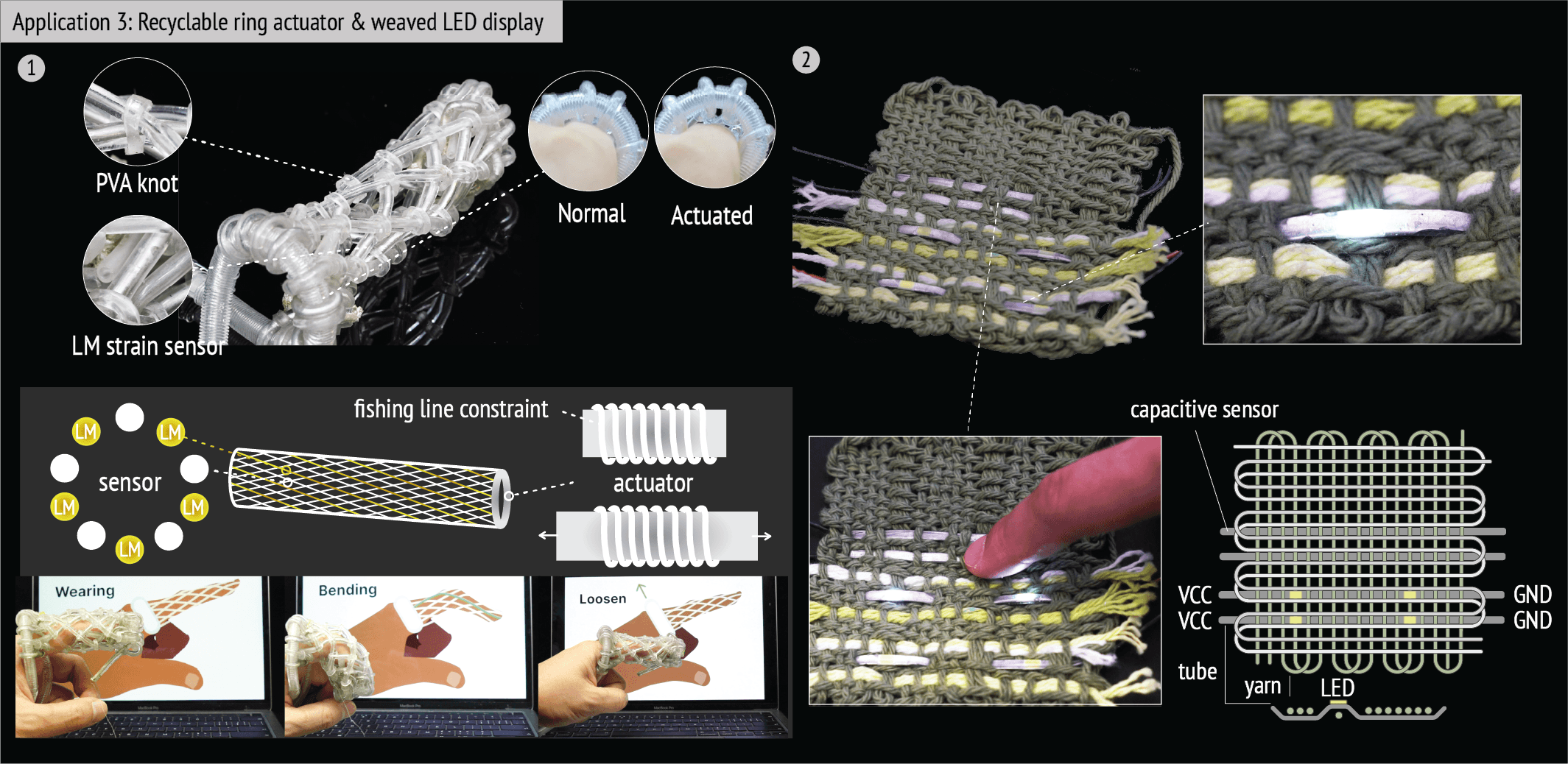}
  \caption{Two tube-based applications. (1) A fully recyclable sensor embedded haptic ring, and (2) An hand-weaved interactive LED.
 }~\label{figure:tubeapplication}
\end{figure*}

For building recyclable tube-based interactive devices, we present two applications. Firstly, we introduce a strain-sensor enabled haptic ring, as shown on the left side of \textcolor{blue}{Figure \ref{figure:tubeapplication}}. This device comprises ten PVA tubes, each has 1.5mm in internal diameter (ID), 2.9 mm in outer diameter (OD), and 10cm in length. These tubes are hand-woven using PVA-based knots for secure connections. Five of these tubes are filled with LM, serving as strain sensors to detect users' finger movements. Additionally, a circular-shaped pneumatic actuator, fabricated from a thicker PVA tube (3mm ID, 6mm OD, and 20cm long), is integrated at the device's tail end. The pneumatic tube's axial movement is restrained with fishing wire, allowing the actuator to expand or contract around the finger. As a result, when a user bends their finger, the ring can automatically get loosened, enabling easy removal (illustrated at the bottom of \textcolor{blue}{Figure \ref{figure:tubeapplication}}).

Secondly, we demonstrate an interactive LED setup, showcased on the right side of Figure \ref{figure:tubeapplication}. This setup is crafted by weaving four LM-filled PVA wires with conventional cotton yarns. Among these, two wires function as capacitive sensors, which we simply implemented the self-capacitance mechanism and the Arduino CapacitiveSensor library is used. The remaining two LM-filled PVA wires serve as connectors for two SMD LEDs, where two squared PVA tubes are cast to fit the size of the LEDs. This configuration allows for the individual activation and deactivation of each LED by interacting with the corresponding capacitive sensing wire.

\section{Recycling Process}
We have introduced Recy-ctronics for crafting interactive devices across varying form factors. These devices not only offer diverse interaction capabilities but are also designed with full recyclability in mind. The recycling process is the same for all three form factors—sheet, foam, and tube, while recycling rates, duration, and practical viability might differ among different form factors.

The recycling process follows prior literatures \cite{teng2019liquid, xu2021printable, teng2023fully}, which starts with detaching electronic components if there are any (\textit{e.g.,} SMD, MCU, battery) from the substrates. Any material remnants on the pins of the components can be delicately eliminated using tweezers or by partially immersing them in water. Any components can be further dried out to prevent electrical failure under the condition of 80-90°C for one day in a laboratory oven. Then the entire device undergoes immersion in water until complete dissolution is achieved. Throughout this process, the dissolved device is subjected to thorough rinsing with fresh water, a step conducted 2-3 times to ensure the complete dissolution of PVA. The resulting PVA-enriched water is then collected, following the separation of the PVA-rich water from the residual LM fraction, approximately 5ml solution of 1mol\% NaOH is added to the LM. The quantity of NaOH solution is contingent upon the device's size. This NaOH serves the role of removing the oxidized layer on the LM, a thin Ga${2}$O${3}$ shell, thereby enabling the seamless recycling of the LM and ensuring high recycling efficiency. With the removal of the oxidized layer and the transformation of the LM into a spherical form, one can retrieve the LM by using a syringe. 

\begin{figure*}[h]
  \centering
  \includegraphics[width=1\linewidth]{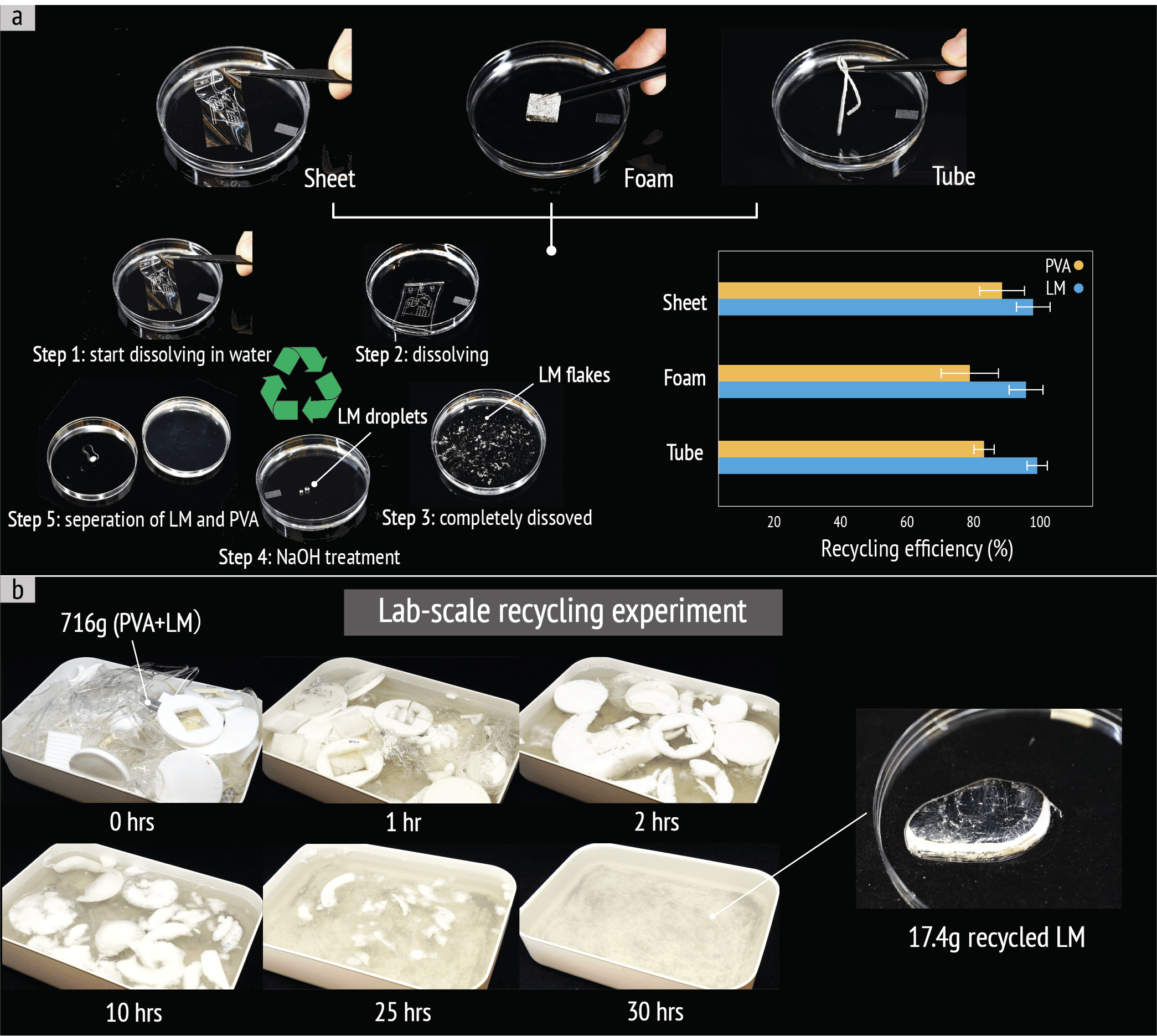}
  \caption{Recycling process. (a) All three form factors-sheet, foam and tube will undergo immersion in the water until complete dissolution. Then the LM and PVA will be separately collected to conclude the recycling process. (b) Lab-scale recycling experiment over two months. 
 }~\label{figure:recycletest}
\end{figure*}

To assess the recycling performance across three distinct form factors, we carried out comprehensive recycling testing. This encompassed recycling an LED pattern on a thin sheet, a cylindrical conductive foam, and a 4cm pipe. The recycling process was executed with six samples for each design, documenting and presenting the pre- and post-recycling weights of PVA and LM. As shown \textcolor{blue}{Figure \ref{figure:recycletest}}, all three sample groups achieved a high recycling rate over 95\% on average for LM, where we have achieved 97.5\%, 96.1\% and 98.6\% for sheet, foam and tube respectively. Also the loss of PVA during the recycling process is higher, where we obtained 87.4\%, 78.1\% and 83.2\% for sheet, foam and tube respectively. Among all the three types of interactive devices, the tube group displayed the highest recycling rate for LM. This lower recycling rate for LM can potentially come from the foam's highly porosity structure, which is mostly covered by LM, inducing more oxidized area compared to sheets and tubes.

Beyond the individual device recycling test phase, we expanded our scope to include a batch of devices accumulated over two months, leading to the recovery of 17.4 grams of LM, as detailed in \textcolor{blue}{Figure \ref{figure:recycletest}} b. This batch consisted of various device types (\textit{e.g.,} sheets, foams, tubes) and miscellaneous non-device waste, totaling 716 grams. Within approximately 30 hours, all materials dissolved in water, which is later rinsed three times to transport the PVA-rich water from the LM remnants. It is worth noting that we paid more attention to recycling LM for this experiment and saved the PVA solutions in containers for later prototyping use.

This lab-scale recycling initiative revealed several insights: (1) \textbf{Complexity in Bulk Processing}: Recycling a multitude of devices introduces complexities that we did not encounter in single-device recycling. The diverse mixture resulted in the emergence of unexpected contaminants such as silver epoxy particles, paper scraps, and wires, requiring further filtration through a 200 mesh filter (around 75 microns). (2) \textbf{Necessity for Improved Sorting}: The heterogeneous nature of the waste highlighted the need for a better sorting system. Varied material compositions with different ratios of glycerin, PVA, and de-ionized water, suggest pre-recycling categorization could stabilize the material composition of recycled outputs. (3) \textbf{Observations on LM Recycling}: The bulk recycling process yielded a higher presence of oxidized substances and an increase in the resistance of the recycled LM, with a sheet resistance of 0.019$\pm$0.011 $\Omega/sq$ (20 pieces of 40cm by 0.5cm samples), compared to the original LM's 0.013$\pm$0.002 $\Omega/sq$, which might stem from the extended environmental exposure during waste accumulation. (4) \textbf{Challenges in Dissolving PVA}: We have observed that the dissolving process dramatically slowed down after several hours, and we had to add more water to the dissolving tank. Also, approaching the end of the dissolving stage, we picked up some remained PVA fragments and separately dissolved them in a beaker with hot water. Overall, this experience with recycling accumulated waste underlines the method's simplicity and efficiency, making it particularly suitable for maker environments.

\begin{figure*}[h]
  \centering
  \includegraphics[width=1\linewidth]{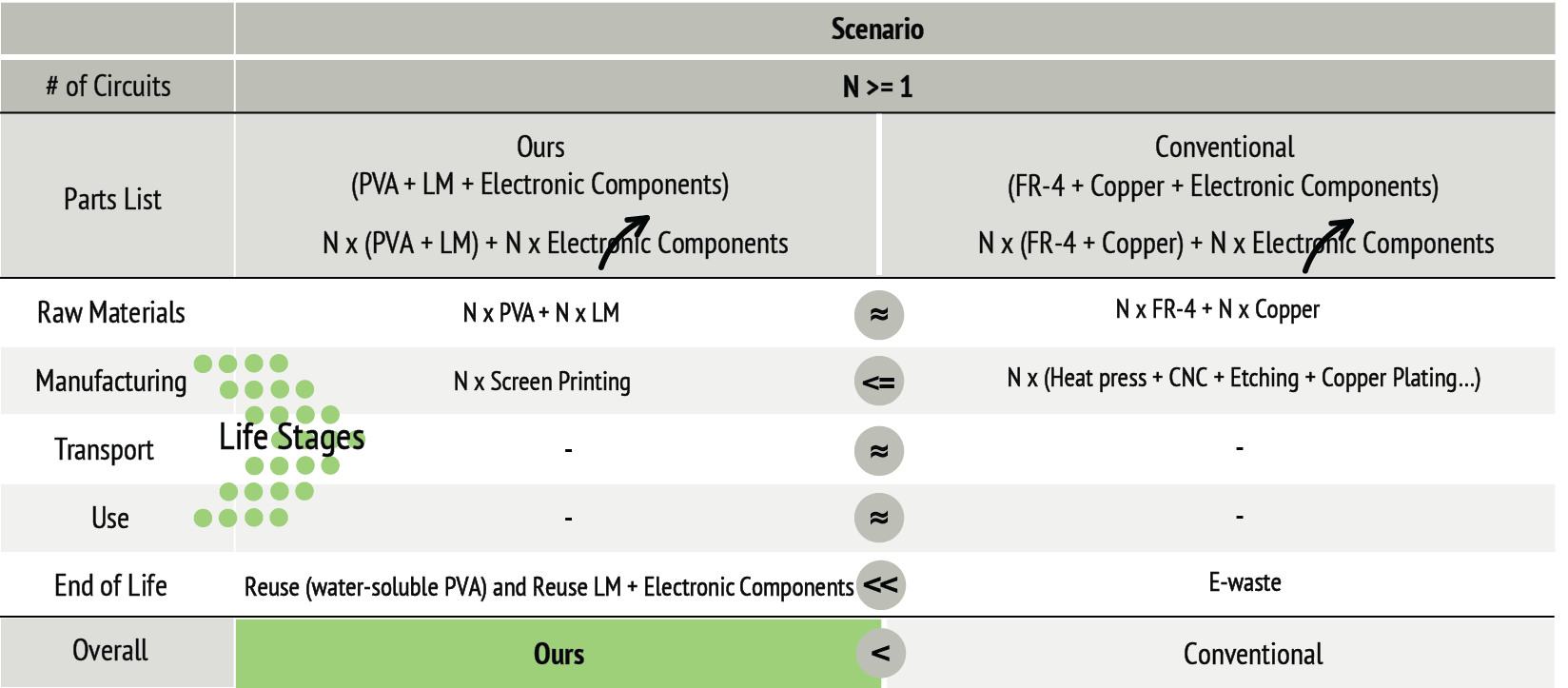}
  \caption{Comparative environmental impact analysis of Recy-ctronics sheet and conventional FR4 PCB.
 }~\label{figure:LCA}
\end{figure*}

\section{Comparative Life Cycle Assessment}

We undertake an environmental impact comparison  (\textcolor{blue}{Figure ~\ref{figure:LCA}}) to juxtapose Recy-ctronics against conventional rigid PCB-based scenario using the case of the recyclable sheet-based proximity sensor presented in \textcolor{blue}{Figure ~\ref{figure:RFID}}. This comparison utilizes the recent concept of a comparative life-cycle assessment (LCA) \cite{zhang_deltalca_2024} to underscore the environmental benefits of Recy-ctronics. FR-4, the most common PCB substrate, serves as the baseline for comparison. We assume identical dimensions to our substrate, i.e., 45mm x 55mm, and a standard thickness of 1.6 mm per dielectric layer. Various common LCA boundaries exist, including ``cradle-to-gate'' assessments,  which encompass raw material extraction through product manufacturing, and ``cradle-to-grave'', which extends to transport to the consumer, usage throughout the product's lifetime, and end-of-life disposal. Given our focus on end-of-life recycling, we adopt a ``cradle-to-grave'' analysis as the primary scope of this comparison study.

In the initial phase of conducting comparative LCA, we compile a parts inventory of the recyclable sheet-based proximity sensor circuit and an FR-4 printed circuit board assembly (PCBA). As demonstrated in the previous section, the Recy-ctronics sheet effectively supports the standard MCU and sensor operation. Consequently, electronic components across both sensors can be canceled out in this environmental impact comparison as they are mostly identical. Thus, the sole distinction between Recy-ctronics sheet and conventional FR-4 circuitry in the raw material phase lies in the substrate, and conductive traces on the circuit board, which are PVA and LM for Recy-ctronics, and copper-clad FR-4, respectively. The environment impact of these parts is approximately equal in the raw material stages.

The first step of comparative LCA is to create a parts inventory of the recyclable sheet-based proximity sensor and an FR-4 sensor, we have shown in the previous section that Recy-ctronics sheet can support the successful operation of a standard MCU and sensor, so the electronic components across could be canceled out in this environmental impact comparison as they are identical. Therefore, the only difference between Recy-ctronics sheet and a conventional FR-4 circuit is the substrate and the conductive traces on the circuit board, and their corresponding manufacturing and end-of-life stage. We qualify them as PVA and LM for Recy-ctronics, and copper-clad FR-4, they are comparable in the raw material stages.

In the manufacturing stage, Recy-ctronics and conventional PCBs each possess pros and cons. For instance, Recy-ctronics employs screen printing to pattern circuits, a method that is faster for low-quantity manufacturing and requires less expensive equipment and ecosystem. However, it lacks scalability compared to the conventional heat press and etching process, which can produce large quantities of parts or multiple boards simultaneously. From an environmental standpoint, screen printing proves more beneficial, as its required equipment consumes significantly less electricity, which contributes almost half of the global warming potential in conventional PCB manufacturing \cite{nassajfar_alternative_2021}, compared to conventional PCB manufacturing equipment in factories \cite{zhang_recyclable_2024}.

We assume similar transportation and use life stages throughout the device's lifespan, as they are the same device. When both devices eventually reach the end-of-life stage. While FR-4 becomes e-waste, Recy-ctronics enables the reuse of its substrate (PVA) and conductive traces (LM), along with electronic components, owing to an accessible recycling process. As discussed before, the recycling process for Recy-ctronics is very straightforward with a high recycling rate which only requires dissolving the device by immersing it in the water, reducing the oxidized layer outside LM and separating the two materials. This entire process is highly accessible without involving complicated tools/steps. For recycling FR-4, the collected PCBs are dismantled and shredded, followed by crushing and grinding to reduce them into finer particles. Subsequent stages involve air separation, screening, and magnetic and electrostatic separation techniques to isolate copper from non-metallic materials. Chemical processes, such as hydrometallurgical methods, are employed to leach out copper, which is then refined for reuse. Meanwhile, the fiberglass component is processed separately to recover fiberglass materials for further applications \cite{xian2021recovery}. 

In summary, Recy-ctronics offers substantial environmental benefits and potentials compared to the conventional, non-recyclable FR-4, particularly at the end-of-life stage.
\section{Discussion, Limitation and Future Work}

Recy-ctronics aims to make interactive devices across a broad spectrum of form factors for various use cases, and with complete recyclability as the core feature in mind. Here, we reflect on several critical factors for Recy-ctronics, including what elements remain non-recyclable in the system, how robust PVA-based devices are, and how we encapsulate the devices.

\subsection{What are not recyclable}
In this work, we introduced both PVA and LM to make fully recyclable electronics with three distinct form factors. Meanwhile, most of the electronic components can also be recycled for reuse, even submerged in the water. One can follow the drying process described previously in the paper by placing chips and components in a glass petri dish
inside a laboratory oven at 80-90°C for one day. The major non-recyclable part in our system is the silver epoxy which is used to connect different components to the LM circuit or wires, and currently we will need to remove the residue for the silver epoxy from both the components and the substrate before we recycle them. Usually silver epoxy contains thermosetting polymers that form cross-linked networks when cured, providing the adhesive with its structural integrity and mechanical strength but making the recycling process challenging. We envision this can be tackled by developing more recyclable type of solder, for example instead of using thermosetting polymers as the base material, we can also use more recyclable materials, such as PVA which is already used in this paper.

\subsection{Robustness for PVA}
Recy-ctronics aims to develop electronics that are transient and fully recyclable, prioritizing sustainability over functionality. However, ensuring the system's reliability and preventing malfunction or early failure are also very important. In this work, PVA is the main water-soluble material and initiates the dissolution process, making the system sensitive to water or even moisture. As also described in previous work, the tunability of PVA's dissolvability can rely on different parameters such as the molecule weight of PVA or different types of PVA \cite{cheng2023functional}. While we believe the water solution nature of PVA made the system not an ideal option for water-related applications (\textit{e.g.,} under-water electronics). Moreover, moisture can significantly impact the mechanical and electrical performance of PVA-based systems. In this study, we explore the use of PVA mixed with glycerin—a simple polyol compound characterized by three hydroxyl (OH) groups. These groups can facilitate the formation of hydrogen bonds with water molecules, including those present in environmental moisture. This results in noticeable changes in our samples' softness or stiffness depending on weather conditions. To mitigate these issues, one strategy involves storing the samples in controlled conditions, such as within airtight bags. Alternatively, in the future, we plan to explore the application of hydrophobic coatings to enhance the water resilience of the PVA substrate, thereby broadening the potential applications of our sustainable electronic systems.

\subsection{Encapsulation and sealant}
Recy-ctronics introduces three types of interactive devices, each requiring different levels of encapsulation. For sheet-type electronics, encapsulation can be achieved by applying an additional thin layer of prepared PVA film. This is done by adhering the PVA film to the substrate layer and slightly spraying the PVA surface with a mist of water. Upon drying, the film forms a tight bond with the PVA layer, completing the encapsulation process. For devices incorporating additional electronic components, the PVA film needs to get laser-cut to create spaces for the components before adhering it to the PVA substrate using the same water mist method. For foam-based electronics, achieving a conformal encapsulation that follows the foam's geometry is essential. To accomplish this, we apply another layer of PVA solution directly onto the foam using a brush. We specifically adopted an elastic formulation for this layer, which is glycerin, PVA, and water in the proportions of 3: 2: 3.5, ensuring that the foam's movement is not restricted. This specific formulation also has a reduced water content to prevent further dissolution of the already cured PVA foam during the brushing step. Regarding the tube-shaped devices, an additional coating layer is not usually needed. However, applying some PVA at the points where the PVA tube connects to wires will further enhance the stability of the wire connections. 
\section{Conclusion}
In this paper, we have introduced a set of readily accessible materials, such as polyvinyl alcohol (PVA) and liquid metal (LM), along with innovative fabrication methods, including screen printing, mechanical stirring, and mold casting. These enable the prototyping of interactive devices in three distinct form factors: sheet, foam, and tube. Each of these device types can undergo a simple immersion in water for an efficient dissolution and recycling process, with an exceptional recycling rate. Furthermore, we proposed a material design strategy that empowers users to tailor the mechanical properties of these devices, ranging from rigid to flexible and stretchable. We hope Recy-ctronics can enable a broad spectrum of users to create versatile interactive devices with sustainability at the forefront, advocating for an environmentally conscious approach prioritizing recyclability over functionality.

\bibliographystyle{ACM-Reference-Format}
\bibliography{reference}

\end{document}